\documentstyle[prl,aps,multicol]{revtex}
\input{epsf}
\begin{document}
\draft
\title{Sensitivity of Quantum Motion for Classically Chaotic Systems}
\author{Giuliano Benenti$^{(a)}$ and Giulio Casati$^{(a,b)}$}  
\address{$^{(a)}$International Center for the Study of Dynamical 
Systems, Universit\`a degli Studi dell'Insubria and} 
\address{Istituto Nazionale per la Fisica della Materia, 
Unit\`a di Como, Via Valleggio 11, 22100 Como, Italy}   
\address{$^{(b)}$Istituto Nazionale di Fisica Nucleare, 
Sezione di Milano, Via Celoria 16, 20133 Milano, Italy}   
\date{\today}
\maketitle

\begin{abstract} 
We discuss the behavior of fidelity for a classically chaotic quantum system 
in the metallic regime. We show the existence of a critical value of the 
perturbation below which the exponential decay of fidelity is determined by 
the width of the Breit-Wigner distribution and above which the quantum decay 
follows the classical one which is ruled by the Lyapunov exponent. The 
independence of the decay {\it rate} from the perturbation strength 
derives from the similarity of the quantum and classical relaxation process 
inside the Heisenberg time scale. 
\end{abstract} 

\begin{multicols}{2}
\narrowtext

Quantum chaos namely the attempt to understand classical dynamical chaos in 
terms of quantum mechanics has lead to a much better understanding of some 
properties of quantum motion which go beyond simple integrable models and 
perturbative treatments. A simple property of quantum conservative 
Hamiltonian systems with a finite number of particles, 
namely discrete spectrum, has been at the origin of some difficulties. 
Indeed in the classical ergodic 
theory discrete spectrum together with linear local instability of motion 
is a typical feature of integrable systems while chaotic systems are 
characterized by continuous spectrum and exponential local instability. 
This fact has questioned the possibility of dynamical 
chaos in quantum mechanics. On the other hand the correspondence principle 
requires transition to classical mechanics of all properties including 
dynamical chaos. As discussed in several occasions
\cite{Chirikov1} this apparent contradiction is resolved by 
taking into account that a sharp distinction between discrete and 
continuous spectrum becomes meaningful only in the limit $t\to\infty$. 
For finite times, there exist different time scales below which the 
quantum motion can display chaotic properties like the corresponding 
classical one. These time scales tend to infinite as the effective 
Planck constant $\hbar_{\rm eff}\to 0$.
Two time scales are of particular importance: the random or the Ehrenfest 
time scale $t_r$ and the relaxation or the Heisenberg time scale 
$t_R$. For $t < t_r$ the quantum motion is exponentially unstable like 
the classical one while the quantum relaxation process takes place during 
the time $t < t_R$. Since typically $t_r << t_R$, the quantum relaxation 
process takes place in the absence of exponential instability. 
A clear illustration of this peculiar feature of quantum motion 
is shown in \cite{cas86}. It should be remarked that this lack of 
exponential instability does not prevent exponential decay of 
dynamical quantities like correlation functions or survival probability  
\cite{maspero}.

Recently the problem of the stability of quantum motion has attracted 
a great interest, also in relation to the field of quantum computation.  
A quantity of central importance which has been on the focus of many studies 
\cite{peres,jalabert,cucchietti,beenakker,tomsovic,prose,flambaum,zurek,cohen} 
is the so-called fidelity $f(t)$, which measures the accuracy to which a 
quantum state can be recovered by inverting, at time $t$, the dynamics 
with a perturbed Hamiltonian:
\begin{equation}
f(t)=|\langle\psi|e^{i\hat{H}t}e^{-i\hat{H}_0t}|\psi\rangle|^2. 
\label{fidelity}
\end{equation}
Here $\psi$ is the initial state which evolves for a time $t$ with the 
Hamiltonian $\hat{H}_0$ while $\hat{H}=\hat{H}_0+\hat{V}$ 
is the perturbed Hamiltonian. 
The analysis of this quantity has shown that, under some 
restrictions, the decay of $f(t)$ is exponential with a rate given by 
the classical Lyapunov exponent \cite{jalabert}. 
This result appears to be consistent 
with recent experiments on the polarization echoes in nuclear 
magnetic resonance\cite{pasta} and with numerical computations
\cite{cucchietti}.
More recent papers have contributed to clarify different 
complementary aspects of the problem 
\cite{beenakker,prose,flambaum,zurek}, 
including the relation with the local density of states 
\cite{cohen} and the use of semiclassical approach 
\cite{tomsovic}. 
The analysis of this quantity has some delicate aspects concerning 
some attempts to characterize quantum chaos via the classical 
Lyapunov exponent and the role of the above mentioned time scales. 
It is therefore highly desirable to have very accurate numerical 
results and to this end it is necessary to consider simple 
systems which display the generic features of classical and quantum 
chaotic systems and which can be easily treated numerically.  

In this paper we consider the behavior of fidelity for a classically 
chaotic system, in the delocalized regime of quantum ergodicity, in 
which the wave functions have a complex pattern which can be described 
within the framework of random matrix theory. 
We show that the type of decay and its rate depend on the strength of 
the perturbation. In particular, above a critical border, the quantum 
decay mimics the classical one and therefore, up to the relaxation 
time scale, it follows the exponential classical decay, which 
in our case is ruled by the Lyapunov exponent.
The independence of the decay {\it rate} on the perturbation, which 
takes place in this regime, simply reflects the properties of the 
underlying classical motion. 

We consider the classical sawtooth map: 
\begin{equation}
\overline{n}={n}+k_0(\theta-\pi),
\quad
\overline{\theta}=\theta+T\overline{n},
\label{clmap}
\end{equation}
where $(n,\theta)$ are conjugated action-angle variables
($0\le \theta <2\pi$), and the bars denote the variables
after one map iteration. Introducing the rescaled momentum variable
$p=Tn$, one can see that the classical dynamics depends only on
the single parameter $K_0=k_0T$. 
The map (\ref{clmap}) can be studied on
the cylinder [$p\in (-\infty,+\infty)$], which can also be closed
to form a torus of length $2\pi L$, where $L$ is an integer. 
For $K_0>0$ the motion is completely chaotic and diffusive, with 
Lyapunov exponent given by 
$\lambda= \ln [(2+K_0+((2+K_0)^2 -4)^{1/2})/2]$.  
For $K_0>1$, the diffusion coefficient is well 
approximated by the random phase approximation, 
$D\approx (\pi^2/3) K_0^2$. 

The quantum evolution on one map iteration is described
by a unitary operator $\hat{U}_0$ acting on the wave function
$\psi$:
\begin{equation}
\overline{\psi}=\hat{U}_0\psi =
e^{-iT\hat{n}^2/2}
e^{ik_0(\hat{\theta}-\pi)^2/2}\psi,  
\label{qumap}
\end{equation}
where $\hat{n}=-i\partial/\partial\theta$ (we set $\hbar=1$).
We take $-N/2\le n < N/2$, $k_0 = (K_0/2 \pi L) N$,
$T=2 \pi L/N$. 
The classical limit corresponds to $N\to \infty$. 
We note that in this simple quantum model one can
observe important physical phenomena like dynamical
localization and cantori localization \cite{fausto}. 
Our aim is to study the fidelity decay in the delocalized regime 
of quantum ergodicity. Moreover we will consider parameter values 
for which there is no initial transient diffusive behavior, which 
may considerably affect the decay of fidelity. 

In order to compute the fidelity we choose to perturb our system by 
slightly varying the kicking strength, $K =K_0 + \epsilon$, 
with $\epsilon\ll K_0$. 
Correspondingly the perturbed quantum kicking parameter
is $k=k_0 + \sigma$, with $\sigma=\epsilon N /(2 \pi L)$.
Since we want to compare classical and quantum evolution, we 
compute the classical ``fidelity'' $f_c(t)$ in the following way: 
we consider in the phase space a uniform density of points 
inside a strip of area $A=2 \pi \nu$ ($0\leq\theta< 2\pi$,   
$-\nu/2\leq p<\nu/2$).  
We then define $f_c(t)$ as the overlap of the initial area 
$A$ with the area $A'$ obtained by evolving $A$ for $t$ iterations 
of the map (2) and then reversing the evolution for $t$ 
iterations with the perturbed strength $K= K_0+\epsilon$.
In practice, we follow the evolution of $10^6$ trajectories 
uniformly and randomly distributed inside the area $A$ and 
define the fidelity $f_c(t)$ as the percentage of orbits which  
return back to the area $A$ at time $t$, after the above reversing 
procedure.  
The corresponding quantum initial condition is given by a 
uniform mixture of momentum states located inside 
the area $A$. We note that this choice, besides giving 
the correct classical limit when $N\to\infty$, introduces a 
convenient averaging procedure. Moreover, we have checked 
that the same fidelity decay rates are obtained if one 
starts from pure states, like momentum eigenstates or 
coherent states.  

\begin{figure}
\centerline{\epsfxsize=8.cm\epsffile{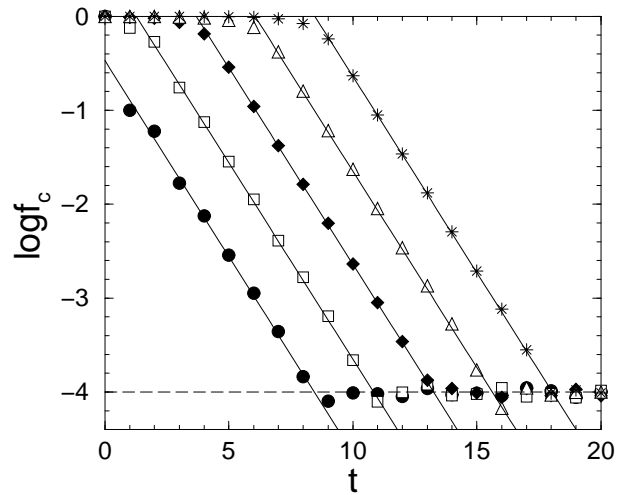}}
\caption{Decay of classical fidelity for the classical sawtooth 
map with $K_0=1$, $L=1$, $\nu=2\pi/10^{4}$, and
perturbation strength $\epsilon=10^{-3}$ (circles), $10^{-4}$
(squares), $10^{-5}$ (diamonds), $10^{-6}$ (triangles), and 
$10^{-7}$ (stars). The straight lines show the decay 
$f_c(t)\propto \exp (-\lambda t)$, with Lyapunov exponent 
$\lambda=0.96$. The dashed line indicates the saturation value 
$f_{c,\infty}=\nu/(2\pi L)=10^{-4}$. Here and in the following 
figures the logarithms are decimal.}
\label{fig1}
\end{figure} 

\begin{figure}
\centerline{\epsfxsize=8.cm\epsffile{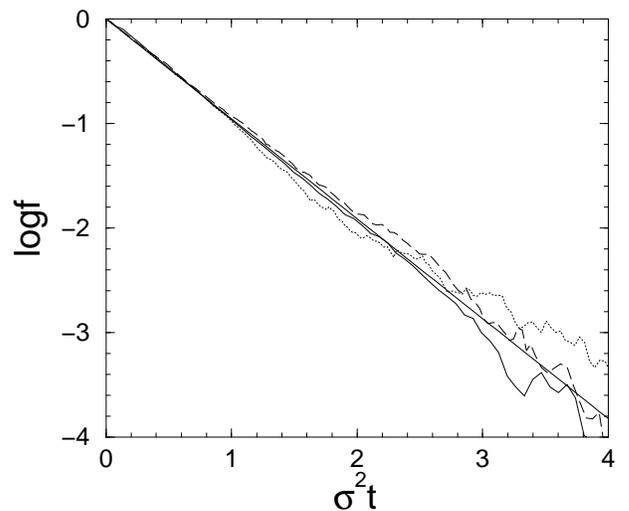}}
\caption{Decay of the fidelity for the quantum sawtooth 
map at $K_0=1$, $L=1$, $\epsilon=5\times 10^{-5}$, 
$N=8192$ (dotted line, $\sigma=0.065$), $16384$ 
(dashed line, $\sigma=0.13$), and $32768$ (solid line, 
$\sigma=0.26$). The straight line gives the decay 
$f(t)=\exp(-\Gamma t)$, with rate $\Gamma=2.2\sigma^2$. 
As initial state we take a momentum eigenfunction with $n=0$.}
\label{fig2}
\end{figure} 

The behavior of the classical fidelity is shown in Fig.1, 
for $K_0=1$, $L=1$, and different values of the perturbation strength 
$\epsilon$. In this particular regime, characterized by   
(i) uniform local exponential instability and (ii) absence of diffusive 
regime, the fidelity decay is ruled by the Lyapunov exponent 
$\lambda$.  
The exponential decay starts after an initial transient time 
proportional to $\ln  (\nu/\epsilon)$, which is required to 
amplify the perturbation up to the scale $\nu$ \cite{eps}. 

\begin{figure}
\centerline{\epsfxsize=8.cm\epsffile{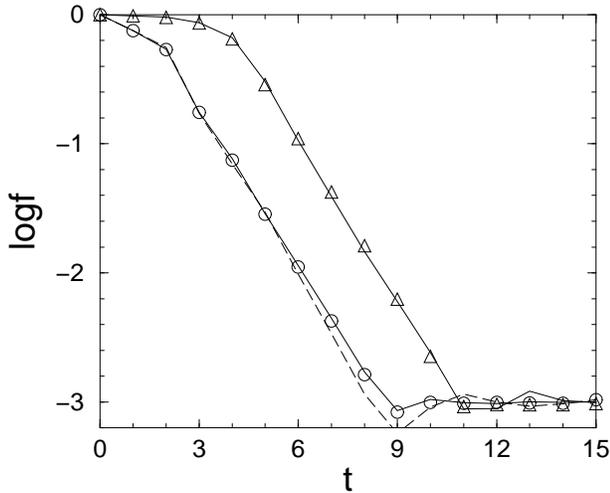}}
\caption{Classical and quantum fidelity decay for 
$K_0=1$, $L=1$, $\nu=2\pi/10^3$. Left curves: $\epsilon=10^{-3}$,
$N=16384$ (dashed line, $\sigma=2.61$), $N=131072$ 
(solid line, $\sigma=20.9$), and classical decay (circles). 
Right curve: $\epsilon=10^{-4}$, $N=131072$ (solid line,
$\sigma=2.09$) and classical decay (triangles).}
\label{fig3}
\end{figure} 

\begin{figure}
\centerline{\epsfxsize=8.cm\epsffile{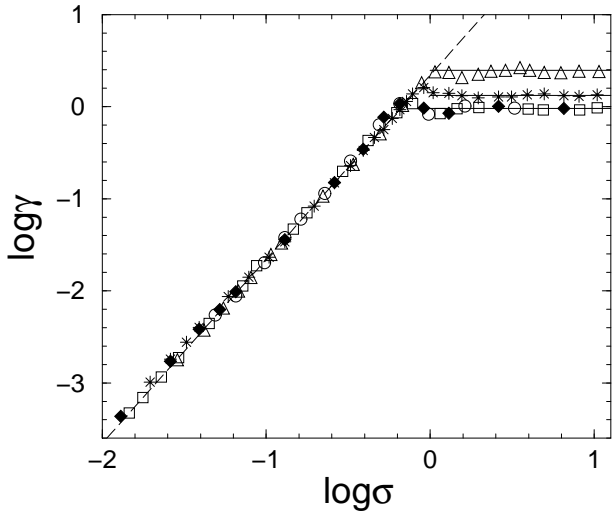}}
\caption{Rate $\gamma$ of the exponential decay for the 
quantum fidelity versus perturbation strength $\sigma$, 
for $K_0=1$, $N=2048$ (circles), $8192$ (diamonds), 
and $65536$ (squares), $K_0=2$, $N=8192$ (stars), 
$K_0=10$, $N=8192$ (triangles). The dashed line gives 
the decay rate $\Gamma=2.2\sigma^2$. The solid lines
show the Lyapunov decay, with rates $\lambda=0.96$
(at $K_0=1$), $1.32$ ($K_0=2$), and $2.48$ ($K_0=10$).}
\label{fig4}
\end{figure} 

The decay of the quantum fidelity is Gaussian below a perturbative 
border \cite{beenakker,tomsovic}. 
This border is given by the value of the 
perturbation at which the typical transition matrix element $U$ 
between quasienergy eigenstates becomes larger than the average 
levels spacing $1/\rho$. For ergodic eigenfunctions, 
$U\sim \sigma/\sqrt{N}$,
while the density of quasienergy states is given by 
$\rho=N/2 \pi$. Therefore the perturbative border is given by
$\sigma_p \approx 1/\sqrt{N}$.
Above this border one typically expects 
an exponential decay of fidelity, with a rate 
$\Gamma =2 \pi\rho U^2 \approx \sigma^2$ given by the  
width of the Breit-Wigner local density of states 
\cite{beenakker}. 
This theoretical prediction is confirmed in Fig.2, which 
shows the decay of quantum fidelity at $\epsilon=5\times 10^{-5}$ 
and different $N$ values, with $\sigma>\sigma_p$. 
The nice scaling behavior of Fig.2 confirms the predicted exponential
decay $f(t) \approx \exp(-C \sigma^2 t)$, with the numerically 
determined constant $C\approx 2.2$.

On the other hand, as stated in the introduction, one expects that 
in the semiclassical regime the quantum motion 
mimics the classical one up to the relaxation time scale which is 
determined by the density of quasienergy eigenstates which 
significantly contibute to the wave function dynamics. 
To  this end it is necessary that the perturbation $\sigma$ is
strong enough to allow the quantum motion to follow, on the average, 
the initial classical decay. In our case this may happen if $\sigma$ 
is large enough to induce transitions at least between nearest 
neighbors momentum states, namely $\sigma > \sigma_c \approx 1$.  
If $\sigma < \sigma_c$, the quantum excitation is unable to follow the 
classical spreading of the initial state. One may also argue in a 
different way: since with our choice of parameters we are in the 
metallic regime, all $N$ quasienergy states are involved in the evolution 
of the unperturbed system. Then the effect of the perturbation on the 
quantum motion can imitate the corresponding classical one only 
if there are no quantum localization effects on the quasienergy
states. This happens when the width of the local density of states 
becomes comparable to the band width, that is $\rho \Gamma \approx N$, 
which again gives the threshold value $\sigma_c\approx 1$. 
We remark that, as discussed in \cite{ccgi}, in the theory 
of Wigner band random matrices the Breit-Wigner regime corresponds  
to a sort of partial perturbative localization. 
The above theoretical estimate is well confirmed by our numerical data 
presented in Figs.3,4. Fig. 3 shows that for $\sigma>1$ the quantum 
fidelity follows closely the classical behavior, namely it decays 
exponentially with the classical rate given by the Lyapunov exponent. 
Fig. 4 shows the decay rate $\gamma$ as a function 
of the perturbation strength $\sigma$. It is clearly seen that for 
$\sigma<1$ the decay rate is proportional to $\sigma^2$, that is to the 
width of the Breit-Wigner. Therefore $\sigma_c \approx 1$ is a critical 
value which separates two distinct regimes: a pure quantum 
perturbation dependent regime, and a semiclassical  
regime. We note that the perturbation $\sigma$ depends on both $N$ 
and $\epsilon$. For $\sigma >1$, the decay rate does not change by increasing 
$N$ at fixed $\epsilon$, since by doing this we merely increase the 
Heisenberg time. On the other hand, if we increase $\epsilon$ 
at fixed $N$ (provided that the perturbation remains classically 
small, i.e. $\epsilon << K_0$) the decay rate also does 
not change, since the exponential amplification of the perturbation is 
controlled by the parameter $K \approx K_0$. In both cases the 
{\it decay rate} of fidelity is perturbation independent. 
This is a property of the classical motion which, in the semiclassical 
regime, is shared by quantum mechanics. 
However, we would like to stress that the decay of fidelity 
remains {\it perturbation dependent}, since the exponential decay starts 
after a time $\propto |\ln\epsilon|$ (see Figs.1,3).   

\begin{figure}
\centerline{\epsfxsize=8.cm\epsffile{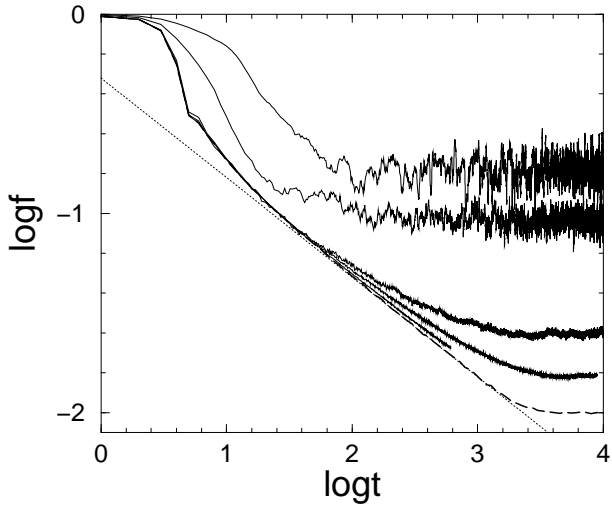}}
\caption{Fidelity decay in the diffusive regime with $L=50$, 
$K_0=1$, $\nu=\pi$, $\epsilon=0.05$. The dashed curve gives
the classical decay, the solid curves the quantum decay: 
from top to bottom $N=1024$ ($\sigma=0.16$), $2048$ 
($\sigma=0.33$), $8192$ ($\sigma=1.30$), $16384$ 
($\sigma=2.61$), and $32768$ ($\sigma=5.22$). 
The straight line indicates the decay 
$f(t)\propto 1/\sqrt{D t}$.}
\label{fig5}
\end{figure} 

For the parameters values of Figs.1-4, the decay of fidelity is exponentially
fast and the saturation value $f_\infty=\nu/(2\pi L)$ 
is reached on times much shorter 
than the Heisenberg time. In order to observe the effect of the Heisenberg 
time scale it is necessary to have a much slower decay of fidelity. 
In Fig.5 we take $K_0=1$ and 
$L=50$, so that we allow for a Gaussian diffusive process in momentum 
space. Because of this, during the diffusion time the fidelity decays 
in the classical case as $1/\sqrt{Dt}$ \cite{nota}. 
Fig.5 shows that for $\sigma>\sigma_c\approx 1$
the quantum decay follows the classical one for larger and larger times 
as $N$ increases, in agreement with the correspondence principle. 
The asymptotic value is $f_\infty=\nu l/(2\pi L)$, where, according to the 
scaling theory of localization, $l=\xi/N=g(x)$, with $x=k^2/N$
\cite{scaling}. 
Here $\xi$ is the actual localization length of the ``sample'' of 
size $N$, while $k^2$ gives the localization length for 
the infinite sample, up to a numerical constant of order $1$.  
The scaling function $g(x)$ is proportional to $x$ for $x\ll 1$ 
and saturates to $1$ for $x\gg 1$. The transition value $x=1$ 
corresponds to $N\approx 10^5$. Moreover, the saturation value 
is approached after a relaxation time $t_l\approx\xi$. 
We stress that in the case of Fig.\ref{fig5} the decay of fidelity 
is controlled by the diffusion coefficient and not by the Lyapunov 
exponent. 
The observation of such regime represents a 
challenge for experiments like spin echoes.  
Further theoretical investigations are also desirable in order to 
understand more clearly the effect of classical diffusion and quantum 
localization on the behavior of fidelity.

In summary, we have shown that the decaying behavior of 
fidelity in a classically chaotic system strongly depends on 
system parameters as well as on the perturbation strength. 
Nevertheless there is a regime in which the decay {\it rate} 
(exponential or power law) is perturbation independent: in 
this regime the  quantum motion simply mimics the 
properties of the underlying classical dynamics. We emphasize
that the quantum to classical correspondence of the average 
behavior is valid until the 
Heisenberg time scale, which is much longer than the Ehrenfest 
time scale associated with the exponential instability of 
quantum motion. 

This work was supported in part by the EC RTN network contract 
HPRN-CT-2000-0156, the NSF under grant No. PHY99-07949, 
the PA INFM ``Quantum transport and classical chaos'', and 
the PRIN ``Caos e localizzazione in meccanica classica e  
quantistica". We gratefully acknowledge the Institute for 
Theoretical Physics, Santa Barbara, California, for the hospitality 
during the initial stage of this work.

\end{multicols}


\begin{thebibliography}{99}
\bibitem{nota} Strictly speaking, $f_c(t)-f_{c,\infty}$ should decay 
exponentially after the diffusive time scale $t_D\approx L^2/D$. 
However this can hardly be seen numerically, as it appears also 
from Fig.\ref{fig5}. 
Actually, for $t>t_D$ the fidelity is already close to its 
asymptotic value $f_{c,\infty}=\nu/(2\pi L)=10^{-2}$.
\end{thebibliography}
\end{document}